\documentclass[twocolumn,prl,aps]{revtex4}
\usepackage{graphicx}

\def\duzomniejsze{<\kern-.7mm<}
\def\duzowieksze{>\kern-.7mm>}

\def\textbf#1{{\bf #1}}
\def\blacksquare{\vrule height 4pt width 4pt depth0pt}
\def\beq{\begin{equation}}
\def\eeq{\end{equation}}
\def\be{\begin{equation}}
\def\ee{\end{equation}}
\def\ben{\begin{eqnarray}}
\def\een{\end{eqnarray}}
\def\beqa{\begin{eqnarray}}
\def\eeqa{\end{eqnarray}}
\def\eea{\end{array}}
\def\bea{\begin{array}}
\newcommand{\bei}{\begin{itemize}}
\newcommand{\eei}{\end{itemize}}
\newcommand{\bee}{\begin{enumerate}}
\newcommand{\eee}{\end{enumerate}}

\def\hcal{{\cal H}}

\def\>{\rangle}
\def\<{\langle}
\def\blacksquare{\vrule height 4pt width 3pt depth2pt}

\def\ot{\otimes}

\def\ccal{{\cal C}}

\def\blacksquare{\vrule height 4pt width 3pt depth2pt}
\def\ot{\otimes}
\def\rh{\rho_{ABA'B'}}
\def\ep{\epsilon}
\def\pbit{pbit}

\def\gammas{pbit}

\def\pdit{pdit}

\def\mainprot{main protocol}

\def\noin{\noindent}
\def\bnoin{\noindent $\bullet \;$}

\newtheorem{proposition}{Proposition}
\newtheorem{theorem}{Theorem}

\newtheorem{observation}{Observation}

\begin{document}
\title{Quantum key distribution based on arbitrarily-weak distillable
entangled states}

\begin{abstract}

States with private correlations but little or no distillable
entanglement were recently reported.  Here, we consider the secure
distribution of such states, i.e., the situation when an adversary
gives two parties such states and they have to verify privacy.  We
present a protocol which enables the parties to extract from such
untrusted states an arbitrarily long and secure key, even though the
amount of distillable entanglement of the untrusted states can be
arbitrarily small.

\end{abstract}

\author{Karol Horodecki$^{(1)}$, Debbie Leung$^{(2)}$, 
Hoi-Kwong Lo$^{(3)}$, Jonathan Oppenheim$^{(4)}$}

\affiliation{$^{(1)}$Department of Mathematics Physics and Computer
Science, University of Gda\'nsk, 80--952 Gda\'nsk, Poland}
\affiliation{$^{(2)}$Institute of Quantum Information, MSC 107-81,
Caltech, Pasadena, California 91125, USA}
\affiliation{$^{(3)}$ Department of Electrical and Computer
Engineering, and Department of Physics, University of Toronto,
Toronto, Ontario M5S 3G4, Canada} 
\affiliation{$^{(4)}$Deptartment of Applied Mathematics and
Theoretical Physics, University of Cambridge U.K.}

\maketitle

\setlength{\parskip}{0.5ex}


Suppose Alice and Bob shared a maximally entangled state, say, an ebit
${1 \over \sqrt{2}} ( | 00\> + | 11\>)$. Clearly, they can generate
a private key directly by measuring their state in the $Z$-basis, 
without any classical post processing.
Are there other types of states with similar key-generating ability?  
Surprisingly, the answer is yes.
Reference \cite{pptkey} gives a necessary and sufficient condition for
a state to generate a key by a direct measurement in the computational
basis -- it must be some {\em twisted} version of a maximally
entangled state called the {\em \pbit} (private bit).

Now suppose Alice and Bob are unsure what state they're sharing.  A
striking feature of entanglement is that, it can be verified and
distilled \cite{BDSW1996}.
Thus, Alice and Bob can first generate near-perfect ebits and then a
private key.  
The best known means to achieve quantum key distribution (QKD) via
noisy, untrusted channels or states is distillation of ebits.
It is then natural to try to go beyond this, by asking whether
noisy and untrusted \gammas s
can similarly be distilled or verified.

The distillation of \gammas s was consider in \cite{pptkey,keyhuge} when
Alice and Bob know they share identical copies of some quantum states.
However, can we achieve QKD with noisy or untrusted {\it \gammas s}? 
In this paper, we provide a positive answer by the explicit
construction of QKD protocols based on noisy \gammas s and by
proving their unconditional security (against the most general attack
allowed by quantum mechanics).
The protocol essentially involves checking bit and phase errors,
with phase errors being checked using a sub-linear number of ebits.  
In the case when an adversary claims to give the parties copies of
ideal private key, which is always distillable, this sub-linear number
of ebits can be obtained by applying an initial distillation protocol 
on some of the key states.
However, there are also states which approximate \gammas s, yet
contain no distillable entanglement \cite{pptkey}.  For these states,
our protocol requires a sub-linear amount of ebits as an extra
resource.  

We will begin with a review of known properties of \gammas s.  We then
introduce the protocol, and prove its security.  Our security proof
also relies on the composability of distillation protocols, and we 
provide a proof in the Ben-Or-Mayers model \cite{Ben-Or-Mayers}. 


%

\noindent {\bf Private states, twisting, and their properties}
 
Suppose Alice and Bob share a quantum state $\rho_{AB}$ and the
eavesdropper Eve has the purification (with her reduced density matrix
denoted by $\rho_E$).
We say that $\rho_{AB}$ contains ideal security if and only if there is 
a local measurement taking it to some ideal ccq state 
\be 
 \rho_{ccq}^{ideal} = \sum_{i}^{d} {1\over d} \, |ii\>\<ii|\ot \rho_E ,
\label{idealccq}
\ee 
signifying that Alice and Bob each has a copy of the key $i$ that is
uncorrelated with Eve.
The class of states containing ideal security in this sense has been
fully characterized in the following way:

{\theorem \cite{pptkey,keyhuge} \label{thm:gamma}
Any state $\rh$ of a Hilbert space
$\hcal_A\otimes\hcal_{A'}\otimes\hcal_B\otimes\hcal_{B'}$
with dimensions $d_A=d$, $d_B=d$, and arbitrary $d_{A'}$, $d_{B'}$,
gives an ideal ccq state after measurement in the computational basis
on the $AB$ subsystem if and only if 
\be
\rh = 
{1\over d} \sum_{i,j=0}^{d-1} |i i\>\<j j|_{AB}\otimes U_i \rho_{A'B'}
U^{\dagger}_j
\label{eq:pdit}
\ee
where $\rho_{A'B'}$ is an arbitrary state of the subsystem $A'B'$ and
the $U_i$'s are arbitrary unitary transformations.
\label{th:pdit}
}

We will refer to a state of the form (\ref{eq:pdit}) as a ``private
state'' or a ``gamma state'' or a ``\pdit'' (and \pbit\ when $d=2$).
Following the convention of \cite{keyhuge}, we will call subsystem
$AB$ the ``key part'' of the pdit and $A'B'$ its ``shield.'' 
These definitions are summarized in Figure 1.  

\begin{figure}[h]
\centering
\includegraphics{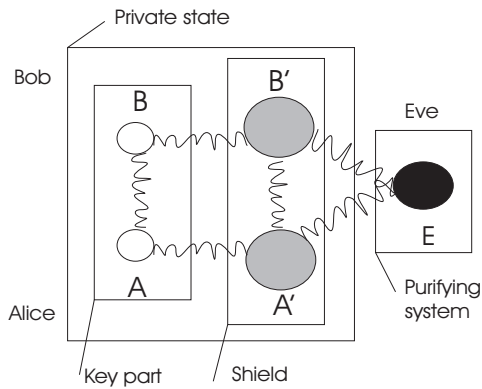}
\caption{ A private state $\rh$ with purifying system $E$.
The ``key part'' ($AB$) after a complete von Neumann measurement gives
an ideal key, which is secure due to the fact that Alice and Bob hold
the ``shield'' part ($A'B'$). \label{fig1}}
\end{figure}

Due to Theorem \ref{thm:gamma}, the distillable key $K_D$ of a quantum
state $\sigma$ can naturally be defined as the maximum ratio of the
logarithm of the dimension $d$ of the output \pdit\ to the number of
copies of $\sigma$ used, and the ratio is maximized over asymptotic
$LOCC$ protocols \cite{pptkey,keyhuge}.

Recall that any private state is a ``twisted'' maximally entangled
state \cite{pptkey,keyhuge}, with the twisting operation defined as
\be
U^{(2)} = \sum_{ij} |ij\>\<ij|_{AB} \ot U^{\dagger}_{ij A'B'} 
 \label{prop:globtwist}
\ee
where $U_{ii}=U_i$ as defined in (\ref{eq:pdit}).  Since twisting is
reversible, we can see this in reverse: any \pdit\ can be turned into
a maximally entangled state on $AB$ and some (global) ancillary state
$\rho_{A'B'}$ on $A'B'$ by a certain twisting operation.  More
formally:

{\observation Consider any private state
$\rh$ of the form (\ref{eq:pdit}) and the twisting defined as in
(\ref{prop:globtwist}).
  $U^{(2)}$ is called a 
``global untwisting'' -- it takes $\rh$ in (\ref{eq:pdit}) into a state
\be
P_+\ot \rho_{A'B'}
\ee
called the basic pdit, where $P_+=\sum_{ij=0}^{d-1}{1\over d}|ii\>\<jj|$
is a maximally entangled state on $AB$ and $\rho_{A'B'}$ is the same
as in (\ref{eq:pdit}). 
The same state change can also result from applying a ``local untwisting'' 
defined as 
\be
U^{(1)} = \sum_{i=0}^{d-1}|i\>\<i|_{B} \ot U^{\dagger}_{iiA'B'} \,.
\label{obs:locuntwist}
\ee
}

Note, that if Bob had access to $A'$ he can transform a private state
into a basic pdit using local untwisting (thus the name ``local'').
The global and local untwistings are respectively subscripted by $(2)$
and $(1)$ (labeling the number of control systems).
Together with the obvious fact that exact teleportation of a system
can be viewed as an identity map on it, we have the following
observation:
{\observation For any state $\rho_{ABA'B'}$, the composition of (i)
teleportation of $A'$ to Bob's side and (ii) local untwisting on
$BB'A'$ commutes with measurement in the computational basis on $AB$.
\label{corol:comm}
}

A final property  of \gammas s to review is as follows:

{\proposition \cite{pawel-strict} For any private state 
$\rho$, $E_D(\rho)\! > \!0$.  
\label{prop:pbitdist}
}

\vspace{0.5ex} 
\noin {\bf Remark} This only holds for exact \gammas s, since one can approximate
\gammas s with bound entangled states.

This concludes our summary for the known results on private states in
the promise scenario in which Alice and Bob know that they share
multiple copies of a certain state.  We now switch to the general
scenario.  We will first describe our main protocol for QKD using
noisy \gammas s, and then establish its unconditional security
against the most general attack by Eve.

\noin {\bf The \mainprot, $M$}

\noin There are six major steps in the \mainprot: 

\bnoin {\it State distribution }

Alice and Bob request $n$ copies of a certain private state
$\gamma_{ABA'B'} \in B(\ccal^d\ot \ccal^d \ot \ccal^{d_{A'}}\ot
\ccal^{d_{B'}})$ given by (\ref{eq:pdit}). We consider the most
general attack where the $\gamma$ states are distributed by Eve.
Therefore Alice and Bob may have an arbitrary joint state over all $n$
systems.  Without loss of generality, we take $d_{A'}\leq d_{B'}$ and
assume compression has already been performed on subsystem $A'$ to
reduce its dimension.

\bnoin {\it Partial distillation}

Alice and Bob randomly choose $k$ out of $n$ systems and run a
distillation protocol that would return $(m\times \log d_{A'}) +t$
\cite{log} ebits if the input were indeed $\gamma^{\ot k}$.  
Alice and Bob estimate the quality of $m\times \log d_{A'}$ of those
untrusted ebits using the other $t$, say, using the Lo-Chau protocol 
\cite{Lo-Chau}.
Here, $t$ is based on a quality parameter $0 < \ep <1$, such that they
abort the protocol with high probability if the fidelity between the
untrusted and ideal ebits is less than $1-\ep$.

\bnoin {\it Random sampling, untwisting, phase-error estimation }

Upon passing the test, Alice and Bob will have $n-k$ systems and $m
\times \log d_{A'}$ distilled ebits.  They pick a random subset of $m$
out of $n-k$ systems, and Alice teleports the $m$ $A'$ subsystems to
Bob using the $m \times \log d_{A'}$ distilled ebits.  To each
teleported $A'$ (together with his local corresponding system $BB'$)
Bob applies the local untwisting $U^{(1)}$ for $\gamma$, as in
(\ref{obs:locuntwist}), to obtain $m$ ``untwisted'' systems.  On the
$m$ ``untwisted'' systems Alice and Bob measure $\sigma_x$ on $A$ and
$B$ and share the results to effect a measurement of $[\sigma_x \ot
\sigma_x ]_{AB}\ot I_{A'B'}$ and estimate the phase-flip error rate
$e_x$.

\bnoin {\it Random sampling and bit-error estimation}

They pick another random subset of $m$ out of $n-k-m$ systems and
measure $\sigma_z$, share their results, and effectively measure
$[\sigma_z \ot \sigma_z ]_{AB}\ot I_{A'B'}$.  This time, they obtain
the bit-flip error rate $e_z$. 

\bnoin {\it  Raw key generation}

If both $e_x$ and $e_z$ are reasonably small, Alice and Bob generate a
raw key from the $n-k-2m$ remaining systems by measuring $[\sigma_z
\ot \sigma_z ]_{AB}\ot I_{A'B'}$ on each of them.
Otherwise, they abort the protocol.  

\bnoin {\it Error correction and privacy amplification}

On the raw key, Alice and Bob perform the two-way Gottesman-Lo
classical error correction and privacy amplification
\cite{Gottesman-Lo} -- repeated concatenation of BXOR and three-qubit
phase code followed by one-way error correction/privacy amplification
(ec/pa) procedure.
\blacksquare

We comment on some aspects of this protocol.
First, Alice and Bob can perform any distillation protocol, even those
assuming tensor power input state $\gamma^{\ot k}$ (e.g.\ the
``hashing'' protocol of \cite{DevetakWinter-hash}).  This is because
having performed such protocol Alice and Bob subsequently check the
quality of the distilled states.  
Second, we do not have to assume that the specific $\gamma_{ABA'B'}$
is distillable -- instead, it is guaranteed by proposition
(\ref{prop:pbitdist}) for all private states.
Third, in the phase-error estimation, the local untwisting operation
can be replaced by any global untwisting.
While these two options are equivalent for perfect private states, 
they are generally different outside of the promise scenario.  
The global untwisting requires the extra teleportation of the $A$
subsystem and thus the distillation of $m \times \log d $ additional
ebits, but can give higher rate than using local untwisting (e.g.\ as
in case of the mixture of two orthogonal private states
\cite{smallkey}).

\noin {\bf Proof of unconditional security of \mainprot\ }

Before stating the proof, we discuss the ideas behind it.  
The unconditional security of $M$ is by reduction to that of the
Lo-Chau protocol \cite{Lo-Chau} based on entanglement purification. 
This reduction is possible because private states are ``twisted
maximally entangled states.''  
Thus, the first step in the proof is to realize that, if Alice and Bob
could (locally) untwist all $n$ systems, Alice and Bob share some
noisy maximally entangled states on the $AB$ subsystems, and standard
techniques \cite{Lo-Chau,Shor-Preskill,Gottesman-Lo} apply so that
the scheme is secure.
The second step is to realize that Alice and Bob do not need to
untwist most of the systems, except for those used in phase error
estimation, and those are indeed untwisted in the \mainprot\ $M$.
This is because the untwisting is followed by the entanglement
purification schemes 
and then measurements
\cite{Lo-Chau,Shor-Preskill,Gottesman-Lo}, a sequence of operations
that can be replaced by measurements followed by classical postprocessing.
But by observation \ref{corol:comm}, the measurements can be done 
before untwisting, which is then unnecessary.  
These replacement are security-preserving, so that we obtain the
desired security of the \mainprot.

For clarity we will first assume Alice and Bob perform errorless
teleportation and local untwisting, and then consider the case when 
these operations are only performed with certain fidelity.

\noin {\it (i) The case of ideal quantum operations}

\bnoin {\it Security of fully untwisted protocol $M_1$ from 
\cite{Gottesman-Lo}}

Let us first consider another protocol $M_1$ that differs from the
\mainprot\ only by an additional step of untwisting (teleporting $A'$
and local untwisting) the $n-k-m$ systems before the measurements in
bit-error estimation and raw key generation.
We now show that $M_1$ is unconditionally secure.
Since Alice and Bob have performed all untwisting operations in
$M_1$, they can trace out the $A'B'$ subsystems, which is equivalent
to giving these subsystems to Eve and can only decrease security. 
Thus, without loss of generality, the input to $M_1$ can be taken to
be 2-qubit noisy maximally entangled states, and results based on
entanglement purification procedures are directly applicable.
In particular, using \cite{Lo-Chau}, if the bit and phase error rates
 are well estimated, the appropriate entanglement purification
 procedure will give a secure key.  The {\it efficient} error
 estimation of \cite{lo-chau-ard} provides good estimate of error
 rates that would have occured if the rest of states were measured along
 the Bell basis.
Thus, after estimating the error rates, Alice and Bob could apply an
appropriate two-way distillation procedure and obtain a secure key by
measuring in bit basis.
Now, \cite{Gottesman-Lo} also states that this can be done
by first measuring in bit basis, and then performing ec/pa, which
gives our $M_1$ protocol.  Since the Gottesman-Lo procedure assures a
secure key, we conclude that $M_1$ is unconditionally secure.

\bnoin {\it Security of \mainprot\ $M$ from that of $M_1$ }

Recall that $M$ and $M_1$ only differ in the additional untwisting on
the systems used in the bit-error estimation and the raw-key
generation steps.  We now show that the extra untwisting is
unnecessary for the security of $M_1$.
Observation \ref{corol:comm} tells us that untwisting commutes with
measurement in the computation basis.  Hence it cannot change
measurement outcomes obtained in the bit-error estimation step and the
raw key generation steps, and thus the values of the estimated
bit-error rate and the raw key.
It follows that untwisting of these $n-k-m$ systems does not effect
the value of the final key and it is unnecessary. 
Thus $M$ differs from $M_1$ only by omitting the necessary untwisting, 
and its security follows from that of $M_1$. 

This ends the proof of unconditional security of the \mainprot\ in case of
ideal operations of teleportation and untwisting.

\noin {\it (ii) The case of imperfect quantum operations}

We now consider the case when Alice and Bob share the maximally entangled 
state and can perform teleportation and local untwisting only up 
to some confidence level. In other word, we assume that
\ben
\|\sigma - P_+ \|_{\rm tr} < \ep \label{eq:dist}\\
\forall_{\rho} \quad \| \Lambda^{noisy}_{te}(\rho)-\Lambda_{te}^{ideal}(\rho)
\| \leq \ep_1
\label{eq:nptel} 
\\
\forall_{\rho} \quad 
\| \Lambda^{noisy}_{untw}(\rho)-\Lambda_{untw}^{ideal}(\rho) \|\leq \ep_2
\label{eq:npltw}
\een
where, as before, $P_+$ is the projector onto a maximally entangled
state (of appropriate dimension), and $\sigma$ is the state
produced by the imperfect distillation, $\Lambda^{ideal}_{te}$ denotes
perfect teleportation of $A'$ and $\Lambda^{noisy}_{te}$ the actual
transformation accomplished by Alice and Bob.  $\ep$,$\ep_1$,$\ep_2$,
are exponential decaying functions in $n$.  Similar notation holds for
the local untwisting operation in (\ref{eq:npltw}).
We have assumed negligible errors in other operations, such as
measurements.

Note that the estimate of the bit error rate is unaffected by the
above errors (\ref{eq:dist})-(\ref{eq:npltw}).  
Now, we show that if the erroneous operations have bounded errors as
described above, the probability is small that they observe a phase
error rate $e_x'$ different from what they would have obtained ($e_x$)
using ideal operations.
This can be proved directly or by using a general composability
result \cite{Ben-Or-Mayers}.  

In essence, the composability result \cite{Ben-Or-Mayers} guarantees the
following in the Ben-Or-Mayers model: Consider a protocol $\pi$ that
uses a certain ideal resource $\kappa$ and achieves security
quantified by a {\em security parameter} $\ep_\pi$ (this quantifies
the level of insecurity, but we will not go into the definition).
Suppose there is a subprotocol $\kappa'$ providing the resource
$\kappa$ with security parameter $\ep_{\kappa'}$.  Then, the protocol
$\pi'$ that uses $\kappa'$ (instead of $\kappa$) will have security
parameter $\ep_{\pi'} \leq \ep_\pi + \ep_{\kappa'}$.

Thus, without loss of generality, we can analyze a variation of the
\mainprot\ that uses ideal ebits instead of $\sigma$ obtained from
imperfect distillation.  If this new protocol is secure, so is the
original one (up to a degradation of $\ep$ in the security parameter).
In particular, Eve could have jointly attacked the imperfect
distillation procedure and subsequent steps in the \mainprot, and the
composability result still applies in the Ben-Or-Mayers model.
It then remains to consider imperfect operations (\ref{eq:nptel}) and
(\ref{eq:npltw}).

Let $\rho_{in}$ be the state of the $n$ systems distributed in the
first step of the \mainprot, $\rho_{out} = \Lambda^{noisy}_{untw}(
\Lambda^{noisy}_{te}(\rho_{in}))$, and $U^{(1)}$ be the ideal local
untwisting defined by $\gamma$.  By the invariance of norm under
unitary rotation and by the triangle inequality we obtain
\be
	\| U^{(1)} \rho_{in} U^{(1)\dagger} - \rho_{out} \|_{\rm tr} 
	\leq \ep_1 + \ep_2.
\ee
The same procedure consisting of measurements and classical
postprocessing is then applied to $U^{(1)} \rho_{in} U^{(1)\dagger}$
in the ideal case, and to $\rho_{out}$ in Alice and Bob's imperfect
protocol, leading to the ideal and actual phase error estimates $e_x$
and $e_x'$.  Since the trace norm can only decrease under this
procedure, the trace distance between the distribution of $e_x$ and
$e_x'$ is at most $\ep_1 + \ep_2$, as we have set out to prove.
This ends the proof of unconditional security of the most general
version of the \mainprot.

\noin {\bf Distilling entanglement versus distilling unconditionally
secure key}

We will comment now on the distilled/distillable entanglement in the
context of our \mainprot. We denote $K_D^{u,{M}}(\gamma)$ as the
amount of 
key obtained in \mainprot\ ($M$) when Alice and Bob demand $n$ copies
of \pdit\ $\gamma$ given that the joint state passes error estimation
step. We consider also the amount of entanglement {\it distilled} in
that protocol denoted as $E_D^{M}(\gamma)$.

\noindent $\bullet$ {\it Distilled entanglement versus distilled secure key}

For the \mainprot\ one has for any \pdit\ $\gamma$:
\be
E_D^{M}(\gamma) \approx 0 \label{eq:E_D} .
\ee
This comes from the logarithmical sample size $s = O(\log d \log n)$
needed to estimate phase error rate in the efficient protocol of
Lo-Chau-Ardehali \cite{lo-chau-ard}.  Thus the amount of distilled
entanglement per input copy approaches zero with increasing $n$. On
the other hand the value of $K_D^{u,{M}}(\gamma)=c$ is nonzero by
definition.

\noindent $\bullet$ {\it Distillable entanglement versus distilled
secure key}

We now compare the distillable entanglement of \pdit\ $\gamma$ with
the distillable unconditionally secure key.  Below we give an example
of the states showing $K_D^{u,M}(\gamma)$ can be {\it arbitrarily}
greater than $E_D(\gamma)$. It is based on the same state for which
one has $K_D(\gamma) > E_D(\gamma)$ \cite{pptkey, keyhuge}.

\noindent {\bf Example}~~
Consider the \pbit\
$\gamma_{ABA'B'} \in B(\ccal ^2 \ot \ccal^2 \ot \ccal^d \ot \ccal^d)$ of
the form \cite{pptkey}:
\be
\gamma_{ABA'B'}= p|\psi_+\>\<\psi_+|\ot \rho_s +
(1-p)|\psi_-\>\<\psi_-|\ot \rho_a
\label{eq:explpbit}
\ee
where $p={1\over 2}(1+{1\over d})$ and $\rho_{s/a}$ are normalized projectors onto
symmetric/antisymmetric subspace. One has for this state
$E_D(\gamma_{0}) \leq \log(1 + {1\over d}) $ \cite{pptkey}.
This leads to the conclusion that there are states for which the
gap between distillable entanglement and distillable unconditionally
secure key is arbitrarily high:
\ben
K_D^{u,{M}}(\gamma_{0}^{\ot \log d}) \geq c\log d \rightarrow_d \,{\infty}\\
E_D(\gamma_{0}^{\ot \log d}) \leq \log d\log(1 + {1\over d})
\rightarrow_d \,{0}
\een
where in the second inequality we have used additivity of
$\log$-negativity measure, which is an upper bound on distillable
entanglement \cite{Vidal-Werner}.

In summary, we introduce protocols for QKD based on noisy \gammas s,
which are a generalization of singlets. We have found that one can
still distill a key in the adversary model even when the distillable
entanglement is made arbitrarily small. Notice that \gammas s are the
most general type of states that can give a secure key. Therefore,
our work generalizes QKD to the most general type of initial states.

A question which arises is whether a truly prepare-and-measure scheme
exists which does not use the teleportation step. One would thus be
able to extract a verifiable secure key from bound entangled states
(i.e. sates which have strictly zero distillable entanglement).  A
protocol for doing this using quantum tomography was given in
\cite{pptkey}, however a security proof was not given.  Such a proof
will be the subject of a future publication.  Finally, we note that in
the case of {\it noisy} \gammas s, the untwisting operation in our
protocol is not known to be optimal (nor proven suboptimal).

\noindent {\bf Acknowledgments}
We thank Matthias Christandl, Daniel Gottesman, Micha\l{} and Pawel Horodecki,
and Andreas Winter  for valuable discussions, 
and the Newton Institute for their hospitality 
while this research was conducted. 
KH and JO acknowledge
EU grants  RESQ (IST-2001-37559),
QUPRODIS (IST-2001-38877) 
and PROSECCO (IST-2001-39227). KH acknowledges support from the 
Polish Ministry of Scientific Research and Information 
Technology under the (solicited) grant no. PBZ-MIN-008/P03/2003. JO
acknowledges support from the Cambridge-MIT Institute and the Newton
Trust. DWL received support from the NSF under
Grant No.\ EIA-0086038, the Tolman
Foundation and the Croucher Foundation.  HKL received support from
NSERC, the CRC Program, CFI, OIT, PREA, and CIPI.


\begin{thebibliography}{14}
\bibitem[{log()}]{log}
\bibinfo{note}{The logarithm is taken to be base 2 in this paper.}
\end{thebibliography}

\end{document}